\documentclass[%
 reprint,
 amsmath,amssymb,
 aps,
]{revtex4-2}

\usepackage{graphicx}
\usepackage{dcolumn}
\usepackage{bm}

\begin{document}

\preprint{APS/123-QED}

\title{Fully Programmable Plasmonic PT-Symmetric Dimer with Epsilon Near Zero and Phase-Change Materials for Integrated Photonics}

\author{Shahab Ramezanpour}
 \email{shahab.ramezanpour@utoronto.ca}
\author{Amr Helmy}%
 \email{a.helmy@utoronto.ca}

\affiliation{
University of Toronto, The Edward S. Rogers Sr. Department of Electrical and Computer Engineering, University of Toronto, 10 King’s College Road, Toronto, Ontario M5S 3G4, Canada
}%

\begin{abstract}
As photonic systems progress toward enhanced miniaturization, dynamic reconfigurability, and improved energy efficiency, a central challenge endures: the accurate and independent control of optical losses and resonant properties on scalable, CMOS-compatible platforms. To address this challenge, we present a hybrid plasmonic dimer that functions in a non-Hermitian regime, capitalizing on the synergistic interplay between Epsilon Near Zero (ENZ) materials and phase-change materials (PCMs) to achieve superior reconfigurability through electrical modulation. Our approach harnesses non-Hermitian physics by precisely modulating the loss differential among coupled modes alongside their resonant frequencies, thereby steering the system to an Exceptional Point (EP) characterized by emergent phenomena and enhanced perturbation sensitivity. By integrating ENZ materials to control dissipation with PCMs to fine-tune resonant frequencies, our structure achieves robust programmability—delivering at least 16 distinct operational states for coupled resonators. This capability supports deep subwavelength confinement and transitions between EP and non-EP regimes, while the inherently low power consumption of ENZ materials and PCMs under deep-subwavelength confinement offers significant advantages even in high-dimensional configurations. We believe that this work outlines a significant route for next-generation programmable photonics, delivering subwavelength confinement, energy-efficient operation, and high-dimensional optical reconfigurability within an integrated, scalable, and manufacturable platform.
\end{abstract}

\maketitle


\section{Introduction}
Programmable building blocks within photonic integrated circuits (PICs),  are essential to manage the increasing complexity of emerging demands on photonics. Moreover, advanced applications in quantum communication and information processing rely on devices capable of controlling the superposition and entanglement of quantum states of light \cite{2}.

A promising platform for adjusting the frequency, amplitude, and phase of modes is the reconfigurable photonic molecule. Photonic molecules (PMs), systems of coupled atom-like optical resonators, exhibit behavior similar to atoms and molecules, generating quantized energy states and supermodes. These manipulations of photonic states and energy can significantly benefit quantum photonics and signal processing. On-chip integration of photonic components, such as modulators, routers, and filters, paves the way for a new generation of photonic-electronic systems with advanced features. Dynamically controlled two-level and multi-level photonic systems could enable novel photonic technologies, such as on-chip frequency-based optical quantum systems, advanced photonic computation concepts, and topological photonics \cite{3,fang2022ultra,4,5}.

An integrated nonlinear dimer offers potential for producing strongly squeezed light free from noise caused by unwanted parasitic nonlinear processes \cite{6}, the formation of self-reinforcing solitary waves \cite{7,8,9,10}, and optical frequency division (OFD) to generate high-quality microwave oscillations \cite{11}.

A structure, which incorporates a low-index spacer between a metal and a high-index material, supports a hybridized plasmonic waveguide mode tightly confined in the spacer, enabling long-range propagation  \cite{12,13,14}. The composite hybrid plasmonic waveguide, comprising a thin layer of low-index material and a thin metallic layer sandwiched between semiconductor layers, demonstrates a record Purcell factor and highly sensitive photodetection \cite{18,19,20,21}. 

A PT-symmetric system, whose Hamiltonian remains invariant under the PT (parity-time) operator, supports real eigenvalues. However, beyond a critical condition, its eigenvalues become complex, signifying broken PT symmetry. The critical point marking the boundary between real and complex eigenvalues is known as the exceptional point (EP) \cite{22}. One significant application of systems operating near the EP is their enhanced sensitivity due to abrupt eigenvalue changes around the EP \cite{23,24}. The EP can be realized in nonlinear and perturbed dimers as a tunable mechanism to compensate for fabrication imperfections and adjust the system to operate near the EP \cite{25,26,ramezanpour2024highly,ramezanpour2024dynamic}. The observation of EPs has so far been limited to wavelength-scaled systems subject to the diffraction limit. Plasmons, the collective oscillations of free electrons coupled to photons, shrink the wavelength of light to electronic and molecular length scales, enabling subwavelength confinement of optical fields. This makes plasmonic systems an ideal platform for realizing exceptional points (EPs) at the nanoscale. By carefully engineering gain and loss in plasmonic metamaterials or metasurfaces, non-Hermitian degeneracies and EP have been demonstrated. These plasmonic EPs mark a critical step toward compact, ultrafast, and highly sensitive photonic devices beyond the diffraction limit \cite{alaeian2014parity,yang2021non,park2020symmetry,22}

There is an increasing demand for a photonic dimer that exhibits high tunability, low power consumption, and a compact footprint, driven by its broad applicability in on-chip programmable integrated photonics. Our dimer employs hybrid plasmonic waveguiding—a concept recently demonstrated and validated—to deliver unmatched tuning energy efficiency. This performance arises from the exceptional confinement of light within a deep sub-wavelength vertical dimension, enabled by hybrid plasmonic modes in combination with tunable materials (as detailed in our previous work such as \cite{19,20,ramezanpour2025exceptional}). By incorporating tunable substances like indium tin oxide (ITO) and phase change materials (PCMs), our design permits the independent and simultaneous modulation of both dissipation and resonance frequency at the nanoscale along the vertical axis. Consequently, the device can exhibit at least 16 distinct tunable states, thereby offering significantly greater dimensionality compared to conventional dimer systems. 

Furthermore, our approach uniquely integrates non-Hermitian physics by precisely adjusting the loss differential between coupled modes, guiding the system toward an exceptional point (EP) characterized by emergent physical phenomena and heightened sensitivity to perturbations. This method contrasts sharply with earlier techniques that depend on gain/loss manipulation through repetitive fabrication, bulk heating, or nonlinear effects, all of which tend to compromise low-power operation and impede integration with photonic circuits. Notably, although recent advances in PT-symmetric systems integrated with electronics have achieved operation at gigahertz frequencies \cite{cao2022fully}, our design functions at 185-200 THz frequency range, ensuring full compatibility with current integrated photonics technology. Previous demonstrations of electrical tuning to an EP—via independent control of dissipation and resonance—often encounter challenges in photonic circuit integration and scalability \cite{park2021accessing,ergoktas2022topological}. In our method, the strategic use of ITO and PCMs effectively decouples these parameters, thereby not only facilitating efficient tuning to the EP but also ensuring compatibility with integrated photonic platforms and offering a scalable solution for advanced optical systems. 
\section{Principle of Operation and Design}
\begin{figure*}
	\centering\includegraphics[width=15cm]{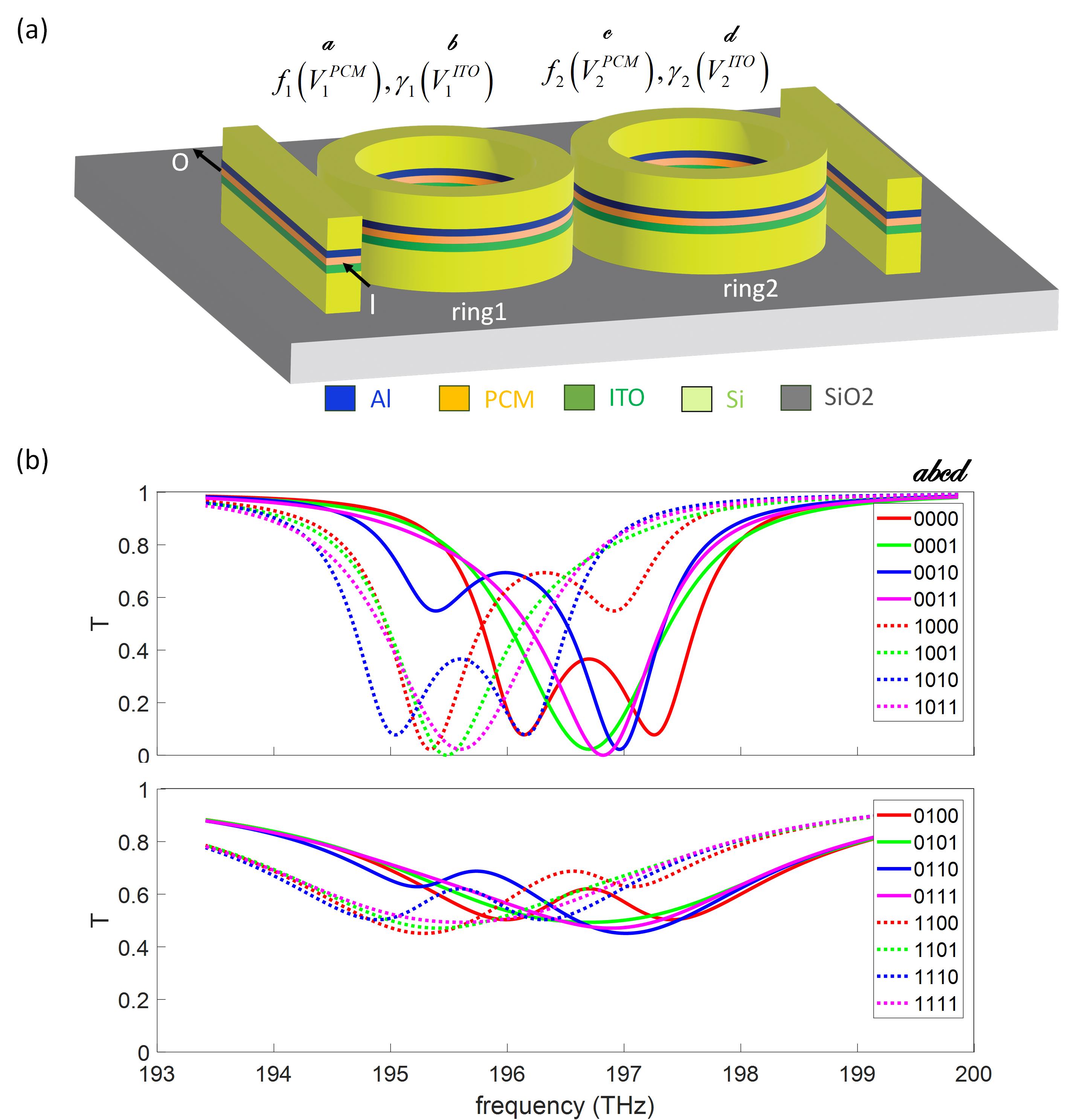}
	\caption{
Programmable Photonic Dimer Enabled by Epsilon-Near-Zero and Phase-Change Materials for High-Dimensional Tunability. (a) The design incorporates CHPW ring resonators constructed from thin layers of low-index material and metal that are embedded within high-index layers. To achieve a fully programmable dimer, the standard low-index materials are replaced with epsilon-near-zero (ENZ) materials and phase-change materials (PCMs). In this configuration, the ENZ materials permit loss modulation through the application of voltage, while the PCMs alter the resonant frequencies by switching between their amorphous and crystalline states. As a result, each resonator within the dimer can function in at least four distinct states, defined by the voltage applied to the ENZ layer and the phase of the PCM. (b) The dimer system’s transmission characteristics reveal at least 16 distinct programmable states, significantly increasing its configuration options and enabling high-dimensional programmability. The digits in the figure correspond to the states of $V_{1}^{PCM},V_{1}^{ITO},V_{2}^{PCM},V_{2}^{ITO}$ representing whether they are on or off, respectively. For instance, $0110$ is associated to \(V_{1}^{PCM}:off,\ V_{1}^{ITO}:on,\ V_{2}^{PCM}:on,\ V_{2}^{ITO}:off\). The transmission is calculated from coupled mode theory (CMT) in which ${{f}_{1,2}}\left( V_{1,2}^{PCM}:off \right)=196.7$ THZ, ${{f}_{1,2}}\left( V_{1,2}^{PCM}:on \right)=195.6$ THZ, ${{\gamma }_{1,2}}\left( V_{1,2}^{ITO}:off \right)=0.4$ THZ, ${{\gamma }_{1,2}}\left( V_{1,2}^{ITO}:on \right)=1.6$ THZ.
}
\end{figure*}
Plasmonic structures enable the confinement of light at the nanoscale, offering deep subwavelength localization. However, their practical utility is often limited by high losses. Hybrid plasmonic waveguides, incorporating a low-index material between metal and high-index material layers, present a promising solution by balancing strong confinement with reduced losses. The Composite Hybrid Plasmonic Waveguide (CHPW) is a specific design featuring thin layers of low-index material and metal sandwiched between high-index layers. This configuration achieves reduced optical losses with nanoscale confinement, high tunability, and low power consumption. The minimized losses result from reduced field interaction with the metallic layer, optimized through careful geometric design. By tailoring this interaction, it is possible to generate both minimized-loss and lossy modes, which can be employed to create passive parity-time (PT)-symmetric systems using coupled minimized-loss and lossy resonators.

In this study, we define the complex frequencies of resonators 1 and 2 as $\Omega_{1,2} = f_{1,2} + i\gamma_{1,2}$, where $f_{1,2}$ and $\gamma_{1,2}$ represent the resonant frequencies and dissipation rates of the respective resonators. For an isolated resonator, its eigenfrequency is inherently identical to its complex frequency. However, in the case of coupled resonators, the eigenfrequencies are determined by both the individual complex frequencies $\Omega_{1,2}$ and the coupling coefficient $\kappa$ between them. It is important to distinguish the complex frequency in this context from complex frequency excitation, where an input signal with an exponentially decaying or growing amplitude is used to introduce virtual gain and loss. Here, the term "frequency" (denoted as $f$) refers exclusively to the excitation frequency. The excitation (bus) waveguide is coupled to the first resonator, and the transmission is evaluated at its output port (through port).

To realize fully programmable dimer functionality, epsilon-near-zero (ENZ) materials and phase-change materials (PCMs) are incorporated. ENZ materials enable loss modulation through applied voltage, while PCMs provide precise control over resonant frequencies by switching between amorphous and crystalline phases, each with distinct refractive indices (Fig. 1(a)).

Here, the ENZ region was realized using an indium tin oxide (ITO) film. Transparent conducting oxides (TCOs) uniquely combine high electrical conductivity with transparency in the visible range \cite{datta2020flexible}, and among them indium tin oxide (ITO) has been a cornerstone of industrial applications for decades \cite{kumar2010race}. In photonic integration, ITO serves as a CMOS compatible epsilon near zero (ENZ) medium, offering tunable free carrier concentration, relatively low near infrared loss, and mature deposition processes \cite{alam2016large}. The ENZ wavelength of ITO can be adjusted across the telecom band by controlling doping density or applying a static electric field, and low temperature ($<$400 °C) annealing further accelerates carrier recombination, enabling ultrafast optical modulation \cite{ni2020property,ma2015indium}. Compared to alternative ENZ platforms such as TiN, ZrN, aluminum doped zinc oxide (AZO), and gallium doped zinc oxide (GZO), ITO offers an optimal balance of tunability, optical performance, and compatibility with established silicon photonic fabrication workflows \cite{kinsey2019near,fomra2024nonlinear}.

We employed a thin Sb$_2$S$_3$ film as our phase change medium, drawn from the ultralow loss PCM family. Phase-change materials such as Sb$_2$Se$_3$ and Sb$_2$S$_3$ exhibit discrete, pulse-dependent changes in their dielectric constant rather than a continuous bias-induced tuning. Crystallization and amorphization are triggered by voltage pulses with defined amplitude and duration; incomplete (partial) switching yields intermediate states. Notably, the optical losses of these materials remain extremely low at 1.55 $\mu$m \cite{del,yang2023non}. Relevant parameters such as pulse amplitude and duration, reported in recent publications on low-loss PCMs, are discussed in detail in the Discussion section.

Leveraging these materials, each resonator within the dimer can operate under at least four distinct conditions, corresponding to the applied voltage to the ENZ material and the phase state of the PCM. When coupled, the dimer system achieves a minimum of 16 programmable states, significantly expanding the configurational space and enabling high-dimensional programmability (Fig. 1(b)). In Fig. 1(b), for a passively identical coupled resonator system, we determined the variation range of the resonant frequency \( f_{1,2}(V_{1,2}^{PCM}) \) and dissipation rate \( \gamma_{1,2}(V_{1,2}^{ITO}) \) by matching numerical results with coupled mode theory (CMT). These variations are analyzed as functions of the applied voltages on the PCM and ENZ layers of each resonator. Additionally, we evaluated the transmission for 16 distinct states using CMT.  

The states are represented as binary numbers, where the first two digits correspond to the first resonator and the last two digits to the second resonator. Within each pair, the first digit indicates the applied voltage to the PCM layer, while the second digit represents the voltage applied to the ENZ material. For example, the state **1001** corresponds to a configuration where the PCM and ENZ layers of the first resonator are in the "on" and "off" states, respectively, while those of the second resonator are in the "off" and "on" states, respectively.

This programmability is further enhanced by the CHPW structure's intrinsic properties: deep subwavelength confinement and high tunability. ENZ materials ensure strong electric field localization within the low-index region, enabling efficient loss modulation without significantly altering the resonant frequency. Conversely, PCMs offer frequency selectivity with minimal impact on dissipation, enabling independent control of both parameters.

The programmable dimer design provides a versatile platform for advanced photonic applications, including reconfigurable photonic circuits, optically controlled logic gates, and high-dimensional quantum state engineering. By combining the unique properties of ENZ and PCM materials, the design achieves a synergistic enhancement of tunability and programmability, paving the way for compact, high-performance devices in integrated photonics.

Figure 2 illustrates the eigenvalue analysis of a CHPW ring resonator coupled to a CHPW waveguide for two whispering gallery modes (WGMs) with angular momentum numbers $m=9$ and $m=10$. The analysis explores variations in the resonator's outer radius and the thickness of the top silicon layer. The CHPW structure uses silicon (Si) and silicon dioxide (SiO$_2$) as high- and low-index materials, respectively, with an aluminum (Al) metallic layer (Fig. 2(a)). The SiO$_2$ and Al layers are 20 nm and 10 nm thick, respectively, while the bottom Si layer is 220 nm thick. The ridge width and the difference between the outer and inner radii of the ring resonator are both 200 nm. The initial geometrical dimensions of the layers in the ring resonator were determined using eigenvalue analysis in COMSOL to achieve the desired resonant frequency and dissipation characteristics. For tunability, the SiO$_2$ thin layer was replaced with tunable materials, followed by FDTD simulations in Lumerical to evaluate the system's performance (Fig. 2(b)). In Fig. 2(b), in the first scheme, the SiO\(_2\) thin layer was replaced with indium tin oxide (ITO) as the active layer and titanium dioxide (TiO\(_2\)) as the passive layer. In the second scheme, both ITO and a phase-change material (PCM) were used as active layers. The first scheme involves passive tuning to the exceptional point (EP), followed by active tuning that shifts the system away from the EP. Conversely, in the second scheme, the system starts in a passively detuned state (outside the EP) and is actively tuned to reach the EP or other specific states.

\begin{figure*}
	\centering\includegraphics[width=15cm]{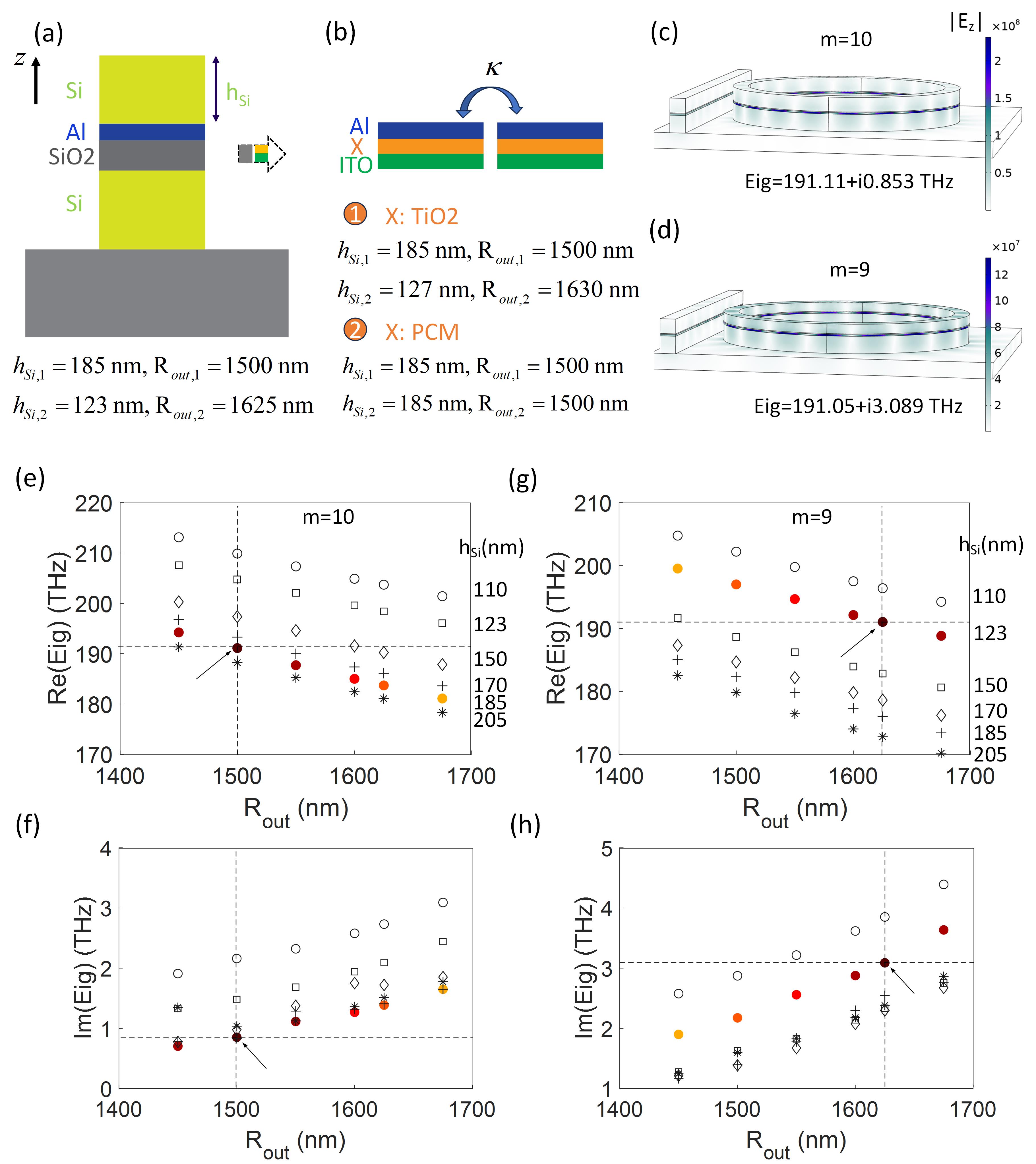}
	\caption{
Eigenvalues of CHPW Ring Resonator
(a) The initial dimensions of the CHPW ring resonator layers were established using eigenvalue analysis in COMSOL to ensure the desired resonant frequency and dissipation characteristics. In this structure, silicon (Si) serves as the high-index material, silicon dioxide (SiO$_2$) as the low-index material, and aluminum (Al) as the metallic layer. The SiO$_2$ and Al layers have thicknesses of 20 nm and 10 nm, respectively, with the silicon layer at the base being 220 nm thick. Additionally, both the ridge width and the radial difference between the outer and inner edges of the ring resonator are set at 200 nm. (b) To enable tunability, the thin SiO$_2$ layer was substituted with tunable materials. In the first configuration, indium tin oxide (ITO) was used, while in the second, a combination of ITO and phase-change material (PCM) was employed. Finite-difference time-domain (FDTD) simulations in Lumerical were then performed to evaluate the system's performance.
(c)-(h) The eigenvalue analysis investigates variations in the resonator’s outer radius $R_{out}$ and the thickness of the top silicon layer $h_{Si}$. A minimized-loss mode (ring 1) is achieved with $h_{Si}= 185$ nm  $R_{out}=1500$, corresponding to a whispering gallery mode (WGM) with $m=10$. A lossy mode (ring 2) occurs with $h_{Si}=123$ nm and $R_{out}=1625$ nm, corresponding to a WGM with $m=9$. The $|E_z|$ component for the minimized-loss and lossy modes is illustrated in (c) and (d), respectively. The real and imaginary parts of the eigenvalues for WGMs with $m=10$ are shown in (e)-(f), and for $m=9$ in (g)-(h).
}
\end{figure*}

To realize an exceptional point (EP), a degeneracy characteristic of non-Hermitian systems and a hallmark of PT-symmetric systems, it is essential to couple modes with identical resonant frequencies while ensuring that the dissipation rate difference equals twice their coupling strength. 

For example, in Fig. 2(e)-Fig 2.(h), the real parts of the eigenfrequencies of two WGMs are matched: ring 1 with $m=10$, a radius of 1500 nm, and $h_{\text{Si}}=185$ nm, and ring 2 with $m=9$, a radius of 1625 nm, and $h_{\text{Si}}=123$ nm (shown by color dots and arrows). The imaginary parts of their eigenfrequencies differ by approximately twice the coupling strength, adjustable by varying the gap between the resonators. The $|E_z|$ field component for both minimized-loss and lossy modes (Fig. 2(c) and Fig. 2(d)) shows that confinement within the low-index thin layer is more pronounced in the minimized-loss mode compared to the lossy mode.

We propose a tunable PT-symmetric structure compatible with integrated photonics, utilizing ENZ and PCM materials to enable electrical tuning of both dissipation and resonant frequencies. This design provides a flexible platform for operating near or away from the EP via an electrically controlled mechanism. Two configurations are explored:
1. **Geometrically Tuned Configuration:** 
The structure is initially tuned close to the EP by optimizing geometric dimensions based on eigenvalue analysis (Fig. 2). ENZ materials are then used to electrically induce loss in the first resonator, transitioning the structure away from the EP by introducing excess loss and making the dissipation in both resonators comparable in the active mode.
2. **Fully Tunable Configuration:** 
This configuration starts with two identical coupled resonators. A PCM is used in the first resonator to tune its resonant frequency, while an ENZ material in the second resonator modulates its dissipation. This dual mechanism enables precise tuning to satisfy both EP conditions: identical resonant frequencies and the requisite dissipation imbalance.

In the first scheme, the low-index material is replaced with 10 nm layers of ITO and TiO2. ITO functions as an ENZ material under applied voltage, providing tunability by inducing loss. TiO$_2$, with a slightly higher refractive index than ITO in the passive mode, ensures the electric field is concentrated within the ITO layer. This enhances device tunability through voltage control.

The inhomogeneous refractive index of ITO under varying voltages is calculated by solving the partial differential equation governing its electro-optic response. Figure 3(a) presents the real and imaginary parts of the ITO refractive index across its thickness, for different applied voltages. The ENZ region, where the real and imaginary parts are nearly equal, initially resides near the ITO/TiO$_2$ interface and shifts slightly further away with increasing voltage.

The electric field intensity peaks within the ENZ region and moves further from the interface as the voltage increases, as shown in Figs. 3(c)–3(f). The effective index of the layered structure, evaluated numerically in Fig. 3(b), varies with applied voltage. The imaginary part of the effective index increases with voltage up to approximately 2 V, beyond which it saturates. The real part shows a slight decrease up to 1 V, followed by a gradual increase to its initial value. These results demonstrate that, at around 2 V, voltage application primarily induces loss while leaving the resonant frequency largely unchanged.

\begin{figure*}
\centering\includegraphics[width=15cm]{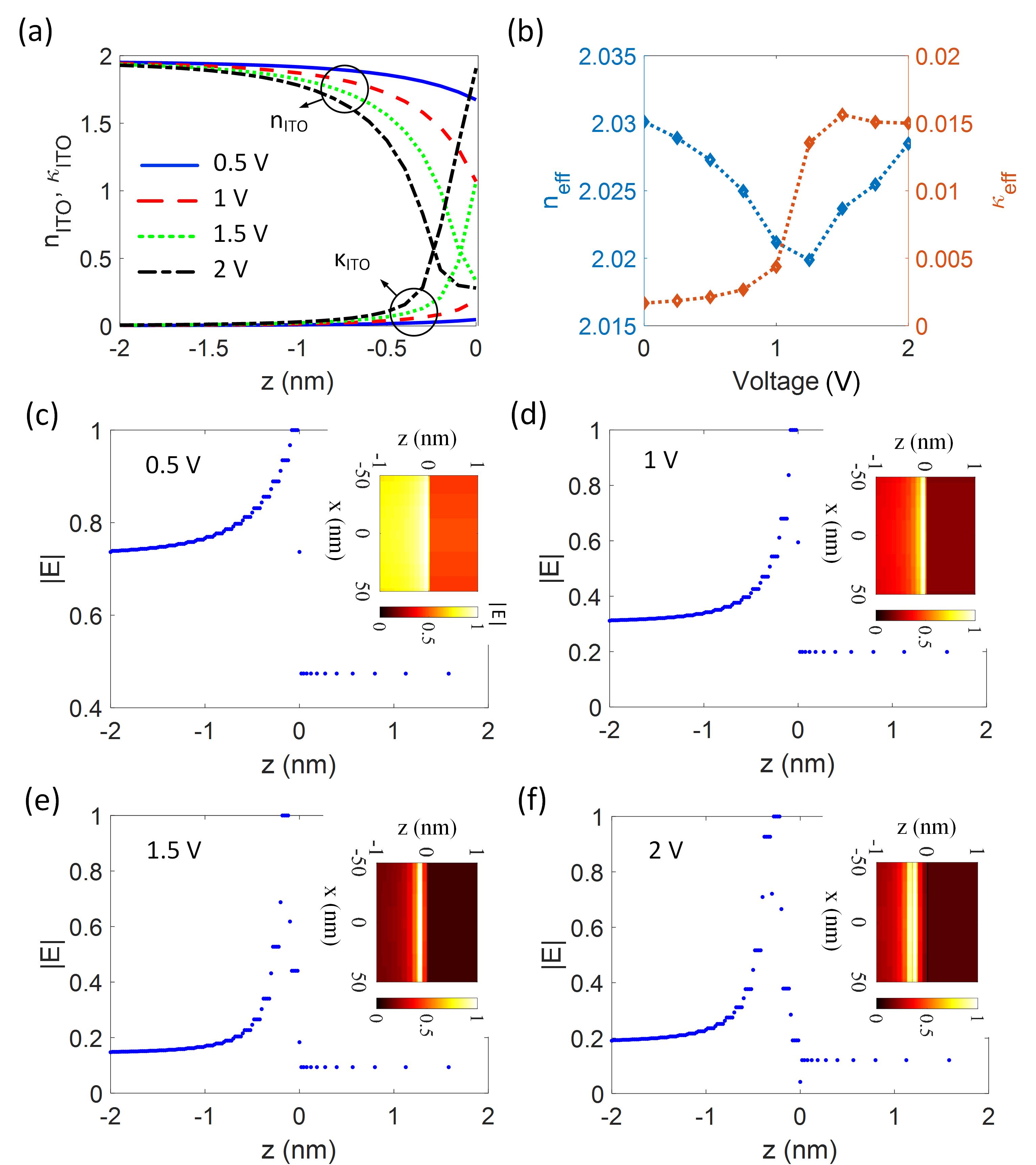}
	\caption{
Inhomogeneous Field Distribution in the ITO Layer
(a) The real and imaginary parts of the ITO refractive index are plotted across its thickness, for varying applied voltages. The epsilon-near-zero (ENZ) region, where the real and imaginary components are nearly equal, initially forms near the ITO/TiO$_2$ interface and shifts further away as the voltage increases. (b) The effective index of the CHPW structure with ITO is shown as a function of the applied voltage. The imaginary part of the effective index ($\kappa_{eff}$) rises with voltage, saturating at approximately 2 V. The real part ($n_{eff}$) slightly decreases up to 1 V before gradually returning to its initial value. These findings indicate that voltage application around 2 V primarily enhances loss without significantly altering the resonant frequency. (c)-(f) The electric field intensity peaks within the ENZ region and shifts further from the interface with increasing voltage. For $V=0.5$ V (c) and $V=1$ V (d), where the ENZ region cannot form, the field distribution remains nearly homogeneous across the ITO layer. However, at higher voltages, such as $V=1.5$ V (e) and $V=2$ V (f), the field intensity becomes concentrated within the ENZ region, which moves further away from the interface.
    }
\end{figure*}
\section{Results}
For the first scheme, Fig. 4(a) presents the transmission spectra of the through port for coupled resonators, where the low-index layer comprises ITO/TiO$_2$ layers, each 10 nm thick, and the metallic thin layer is aluminum (Al) with a thickness of 10 nm. The inner and outer radii of the first ring resonator are 1300 nm and 1500 nm, respectively, with a top silicon layer thickness of 185 nm. For the second ring, the inner and outer radii are 1430 nm and 1630 nm, respectively, with a top silicon layer thickness of 127 nm. The dimensions of the second ring, along with the gap between the resonators, are optimized to approach the exceptional point (EP) conditions for whispering gallery modes (WGMs) with m=10 and m=9, as in Fig. 2.
Using coupled-mode and eigenvalue analysis, the complex frequencies of the resonators are estimated as $f_1+i\gamma_1=197.2+i0.8$ THz and $f_2+i\gamma_2=197.3+i2.2$ THz, with a coupling value $\kappa=0.8$ THz. These values satisfy the EP conditions ($f_1=f_2$ and $\gamma_2-\gamma_1=2\kappa$) (Fig. 4(b)-4(c)).
Applying a voltage to the ITO layer of the first ring modifies its complex frequency to $f_1+i\gamma_1=197.5+i2.2$ THz, indicating a nearly unchanged resonant frequency but an increased imaginary component matching that of the second ring. The complex, resonant eigenfrequency of the second ring and the coupling value remain unchanged.
This significant voltage-induced increase in the imaginary part of the first ring's eigenfrequency results in a pair of (optically) identical lossy/lossy coupled resonators, as evident in the transmission spectra.

\begin{figure*}
	\centering\includegraphics[width=15cm]{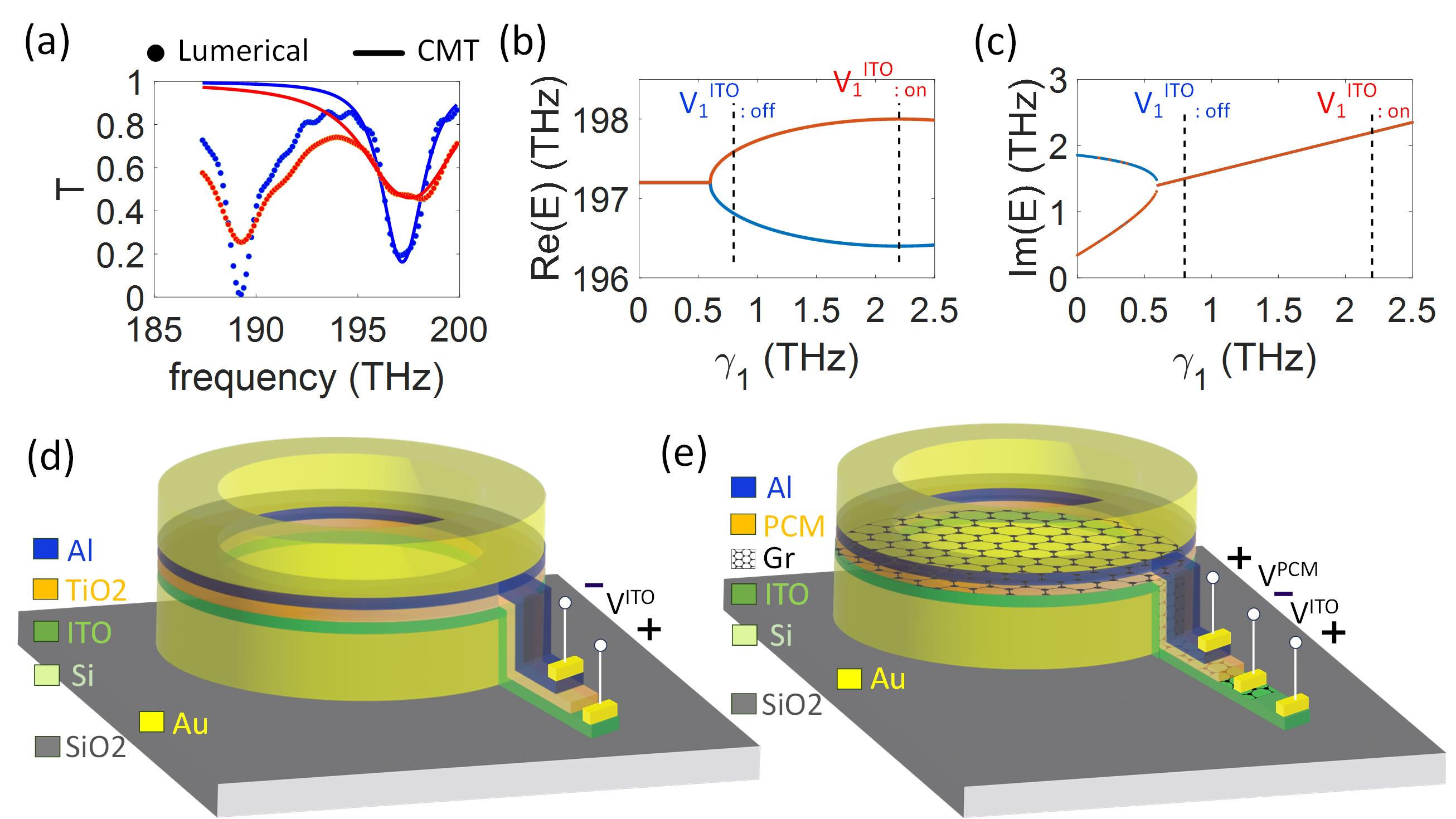}
	\caption{
Scheme 1: Transmission spectra and eigenvalue analysis. (a) Transmission spectra of the through port for coupled resonators are shown. The low-index layer consists of 10 nm thick ITO/TiO$_2$ layers, while the metallic thin layer is aluminum (Al) with a thickness of 10 nm. The ring dimensions are optimized to achieve EP conditions: $\left( {{h}_{Si,1}}=185\ \text{nm},\ {{R}_{out,1}}=1500\ \text{nm} \right)$ and $\left( {{h}_{Si,2}}=127\ \text{nm},\ {{R}_{out,2}}=1630\ \text{nm} \right)$. The dissipation of the first resonator is adjusted by applying a voltage to the ITO layer. The blue and red curves correspond to the ITO layer being voltage-off and voltage-on, respectively. dotted lines represent results from Lumerical simulations, while solid lines show outcomes from coupled-mode theory (CMT) analysis.
(b)-(c) Coupled-mode and eigenvalue analyses are used to estimate the complex frequencies of the resonators as $f_1+i\gamma_1=197.2+i0.8$ THz and $f_2+i\gamma_2=197.3+i2.2$ THz, with a coupling constant $\kappa=0.8$ THz. These values satisfy the EP conditions ($f_1=f_2$ and $\gamma_2-\gamma_1=2\kappa$). When voltage is applied to the ITO layer in the first ring, its complex frequency changes to $f_1+i\gamma_1=197.5+i2.2$ THz, showing that the real part of the frequency remains nearly the same while the imaginary part increases to match that of the second ring. The complex, resonant eigenfrequency of the second ring and the coupling constant remain unaffected. Schematic illustrations of the proposed photonic modulation schemes. (d) Scheme 1: ITO/TiO$_2$/Al layers sandwiched between Si layers, where voltage is applied between the ITO and Al layers to modulate the ITO layer. (e) Scheme 2:  ITO/PCM/Al layers sandwiched between Si layers, with a graphene layer introduced between ITO and PCM as a common ground.Voltage is applied between ITO and graphene to modulate the ITO and between graphene and Al to modulate the PCM.}
\end{figure*}

To modulate the ITO layer in the first scheme (ITO/TiO$_2$/Al), we can apply voltage between the extended Al and ITO layers (Fig. 4(d)). This setup leverages the tunable carrier density of ITO, achieved through electrical modulation, to manipulate its refractive index via the electro-optic effect. This configuration offers precise control over the modulation depth and enables compatibility with high-speed integrated photonic circuits, as the electric field-induced modulation occurs without significantly impacting the optical properties of adjacent layers.
In the second scheme (ITO/PCM/Al), where both the ITO and PCM layers can be modulated, a more versatile approach is required. By introducing a graphene layer between the ITO and PCM, graphene serves as a common ground, enabling separate modulation of the ITO and PCM layers (Fig. 4(e)). Specifically, applying a voltage between the ITO and graphene contacts modulates the ITO's carrier density, while applying voltage between the graphene and Al contacts modulates the PCM layer. This dual-modulation approach is highly advantageous for applications requiring independent or simultaneous control of the two materials, as it allows fine-tuning of their optical and electronic properties within a single device.
To further isolate the ITO and PCM layers and prevent unwanted interference, an additional TiO$_2$ layer can be inserted above the ITO. This layer acts as an insulating barrier, ensuring that the modulation of one layer does not inadvertently affect the other.
In scenarios where contactless methods are preferred to modulate the PCM layer, controllable heating through the Al layer is a viable option. By applying a localized current to the Al layer, heat is generated and transferred to the PCM, inducing a phase transition in the PCM. This phase change can be precisely controlled by adjusting the heating power, allowing for dynamic optical modulation without direct electrical contacts on the PCM layer. Alternatively, laser-induced heating can be employed to selectively modulate the PCM, where a focused laser beam provides the necessary energy to trigger the phase change. These contactless methods ensure compatibility with integrated photonics while avoiding electrical interference with other components in the device.
The combination of these approaches offers a versatile platform for advanced photonic modulation. The electrical tunability of ITO and the phase-change dynamics of PCM enable a wide range of applications, from reconfigurable photonic circuits and optical switching to neuromorphic computing and adaptive optics.

To demonstrate the fully tunable capability of the proposed structure, the low-index layer is configured with ITO and Sb$_2$S$_3$ thin layers, enabling both loss induction and resonant frequency modulation. Sb$_2$S$_3$ acts as a phase-change material, exhibiting refractive indices of 2.7 and 3.3 at a wavelength of 1.55 $\mu$m in its amorphous and crystalline phases, respectively \cite{del}. In Fig. 5, we examine four distinct states. 1: A passively identical coupled resonator system, which is initially outside the exceptional point (EP) $\to$ 2: Applying voltage to the ITO layer of the second resonator tunes the system to the EP $\to$ 3: Introducing voltage to the PCM of the first resonator shifts the system away from the EP $\to$ 4: Adjusting the radius of the first resonator restores the system back to the EP.

In Fig. 5(a)-5(f), we first analyze a pair of identical coupled resonators with inner/outer radii of 1300 nm and 1500 nm and a top silicon thickness of 185 nm. The system is modeled with complex frequencies and coupling values as follows: $f_1+i\gamma_1 = f_2 +i\gamma_2= 189 + i0.6$ THz and $\kappa = 0.8$ THz for the lower band, and $f_1+i\gamma_1 = f_2+i\gamma_2 = 196.7 + i0.4$ THz and $\kappa = 0.63$ THz for the higher band.
When a voltage is applied to the ITO layer of the second ring, the modified complex frequencies for the second ring become $f_2+i\gamma_2 = 189 + i1.4$ THz for the lower band and $f_2+i\gamma_2 = 196.7 + i1.6$ for the higher band, while $f_1+i\gamma_1$ and $\kappa$ remain unchanged. For the higher band, the conditions $\gamma_2-\gamma_1=2\kappa$ and $f_1 = f_2$ are satisfied, achieving the EP condition.
In Fig. 5(g)-5(l), the first resonator is configured in the crystalline phase of the phase-change material, resulting in complex frequency of the first ring of $f_1+i\gamma_1=188.1 + i0.6$ THz for the lower band and $f_1+i\gamma_1 = 195.6 + i0.4$ THz for the upper band, while the complex frequency of the second ring as well as the coupling values are unchanged. To compensate for this frequency shift, the inner/outer radii of the resonator are reduced to 1285 nm and 1485 nm, respectively. This adjustment restores the complex frequency of the first resonator to $f_1+i\gamma_1=189+i0.6$ THz for the lower band and $f_1+i\gamma_1=196.7+i0.4$ THz for the upper band. In the crystalline phase, the upper band also satisfies the EP condition.
\begin{figure*}
	\centering\includegraphics[width=15cm]{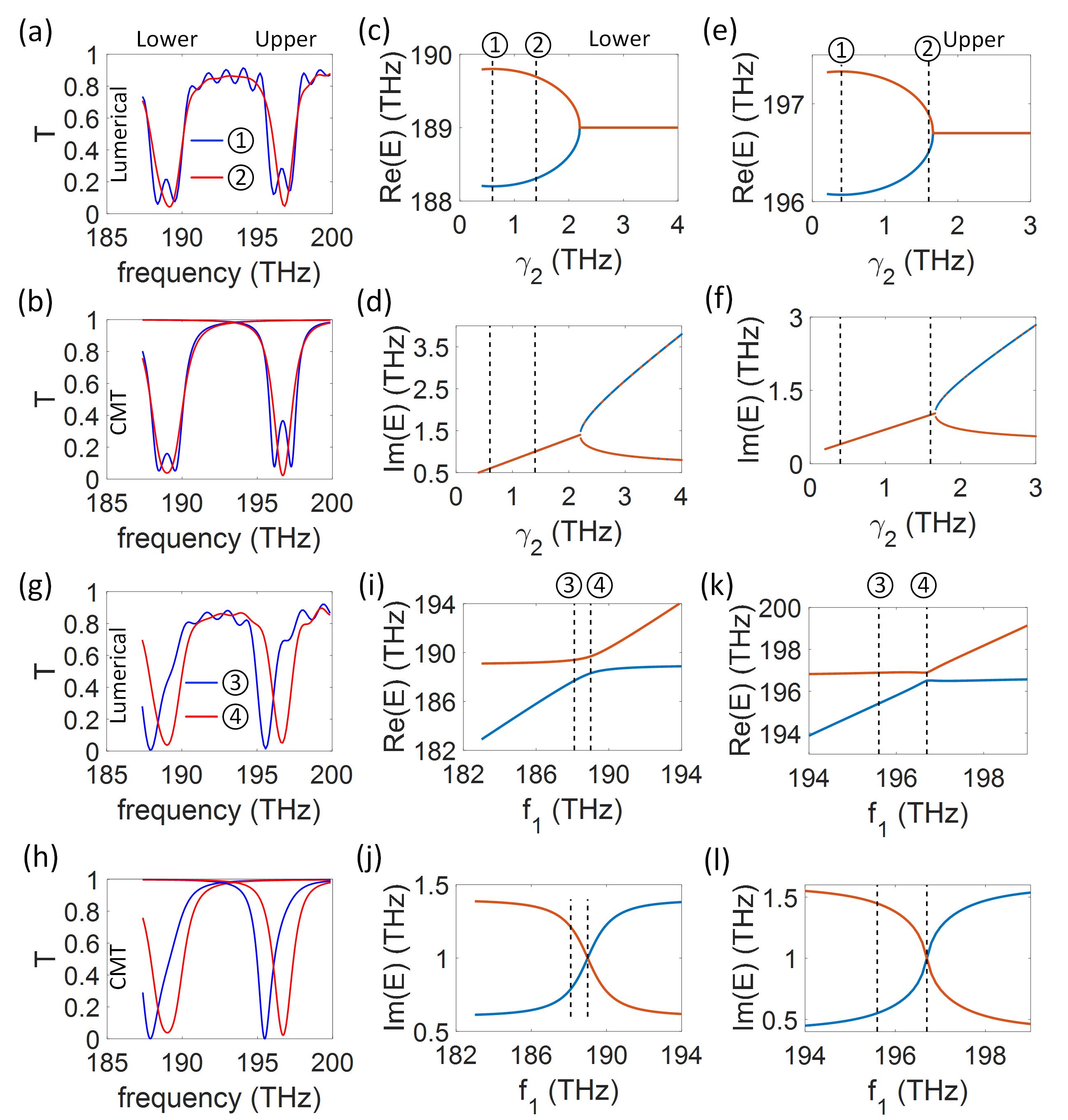}
	\caption{
Scheme 2: Transmission spectra and eigenvalue analysis. (a)-(f) In the first scenario, identical coupled resonators are analyzed. In the second scenario, a voltage is applied to the ITO layer of the second resonator, tuning the system to EP conditions. (g)-(l) In the third scenario, the system is already at EP due to the applied voltage on the second resonator's ITO layer. Additional voltage is then applied to the PCM of the first resonator, causing the system to move away from the EP. Finally, in the fourth scenario, the radius of the first resonator is adjusted, restoring the system back to EP conditions. The complex frequency of the resonators in different states are as following.
State 1 (lower band): $f_1+i\gamma_1 = 189 + i0.6$ THz, $f_2 +i\gamma_2= 189 + i0.6$ THz, $\kappa = 0.8$ THz. State 2 (lower band): $f_1+i\gamma_1 = 189 + i0.6$ THz, $f_2+i\gamma_2 = 189 + i1.4$ THz, $\kappa = 0.8$ THz. State 1 (upper band):  $f_1+i\gamma_1=196.7+ i0.4$ THz, $f_2+i\gamma_2= 196.7+ i0.4$ THz, $\kappa = 0.63$ THz. State 2 (Upper band (EP)):  $f_1+i\gamma_1 = 196.7 + i0.4$, $f_2+i\gamma_2 = 196.7 + i1.6$, $\kappa = 0.63$ THz. 
State 3 (lower band): $f_1+i\gamma_1=188.1 + i0.6$ THz, $f_2 +i\gamma_2= 189 + i1.4$ THz, $\kappa = 0.8$ THz.  State 4 (lower band): $f_1+i\gamma_1=189+i0.6$ THz, $f_2 +i\gamma_2= 189 + i1.4$ THz, $\kappa = 0.8$ THz. State 3 (upper band): $f_1+i\gamma_1 = 195.6 + i0.4$ THz, $f_2+i\gamma_2 = 196.7 + i1.6$ THz,  $\kappa = 0.63$. State 4 (upper band (EP)):  $f_1+i\gamma_1=196.7+i0.4$ THz, $f_2+i\gamma_2 = 196.7 + i1.6$ THz,  $\kappa = 0.63$ THz. 
}
\end{figure*}

These results demonstrate that, in addition to enabling programmable functionality, the PCM and ITO layers can be utilized to compensate for fabrication-induced imperfections. For instance, to achieve tuning to the exceptional point (EP), applying voltage to the PCM layer can counteract an effective radius deviation of up to 15 nm caused by fabrication inconsistencies. Similarly, misalignments in coupling and dissipation between the resonators can be corrected by adjusting the applied voltage to the ITO layer. This capability establishes a self-adjusting platform for mitigating fabrication errors.

In Fig. 5, we demonstrate independent tuning of the ITO layer in the second ring and the phase-change material (PCM, Sb$_2$S$_3$) in the first ring. For the second ring, the electrical contacts can be arranged between the ITO and the Al electrode in a Si/ITO/PCM/Al/Si stack. If the gate bias is kept small (here, below 2 V), the applied field is insufficient to induce resistive switching or significant carrier injection in Sb$_2$S$_3$, so the PCM behaves effectively as a high-resistivity dielectric. In that operating window Sb$_2$S$_3$ may be treated as an insulating layer with a refractive index 2.7 above the ITO, and the ITO carrier density can be modulated capacitively without altering the PCM state.

For tuning the PCM in the first ring, we propose contacts between a monolayer graphene electrode (placed just above the ITO) and Al in a Si/ITO/Gr/PCM/Al/Si stack. A voltage applied between graphene and Al can be used to switch or tune the PCM while leaving the buried ITO largely unaffected, provided the voltage waveform and amplitude are chosen to confine current paths to the PCM layer. To enable simultaneous, independent control of both ITO and PCM within a single ring, one can insert a thin (few-nm) insulating spacer (e.g., TiO$_2$) between ITO and PCM to form Si/ITO/TiO$_2$/Gr/PCM/Al/Si. In that geometry a bias between ITO and graphene primarily gates the ITO, whereas a bias between graphene and Al addresses the PCM, with minimal cross-talk. An alternative stacking is to place the PCM above the Al electrode (Si/ITO/TiO$_2$/Al/PCM/Si): biasing ITO–Al then tunes ITO, while biasing Al–top-Si (or a top electrode on the PCM) addresses the PCM. Note, for the scheme with PCM at top of the Al, electrical switching of the PCM typically modifies both the resonant frequency and dissipation due to the changes of field interaction with Al; achieving a target resonance and linewidth, therefore, requires a coordinated (programmable) tuning scheme that simultaneously adjusts ITO and PCM to compensate these coupled effects. Finally, if long-term endurance or reduced switching energy is required, a tunable doped semiconductor could be used in place of the PCM as an alternative refractive-index tuning mechanism.

These integration options, trade-offs, and control strategies are under active investigation and will be addressed in follow-up work.
See Supplemental Material ~\cite{suppmat2025} for the quantitative comparison (Table S1), full simulation settings, integration guidelines, and the 16-level logic use case (Table S2) (see also references \cite{peng2014parity,cao2022fully,ergoktas2022topological,tiberi2025graphene,yang2023non} therein).

\section{discussion}
Although significant advancements have been achieved in the electronic tuning and dynamic behavior of coupled resonators in photonic atoms and molecules \cite{3,fang2022ultra}, challenges persist in terms of achieving a compact design, high energy efficiency, and a broad range of tunable states. Moreover, within the framework of non-Hermitian physics and the pursuit of accessing exceptional points (EPs), even though electrical modulation of both dissipation and resonance in coupled modes has been demonstrated \cite{park2021accessing,ergoktas2022topological}, issues related to compatibility with integrated photonics and scalability remain unresolved. To overcome these challenges, and in response to the growing demand for on-chip devices that combine high tunability, compactness, and low power consumption, we have exploited the concept of a hybrid plasmonic waveguide platform—recently demonstrated and experimentally verified by our group \cite{19,20}—to design a tunable PT-symmetric dimer. The proposed tunable photonic dimer exemplifies a significant advance in integrated photonics by merging deep subwavelength light confinement, PT-symmetric properties, and hybrid plasmonic design principles. This architecture leverages the unique interplay between Epsilon Near Zero (ENZ) materials and Phase Change Materials (PCM) to unlock a new level of optical programmability. By independently tuning both dissipation and resonance frequency with high precision, the system achieves unprecedented control over light-matter interactions, and establish it as a superior platform for low-power optical modulation. Below, we highlight the key advantages and provide quantitative energy consumption estimations that demonstrate the structure's low-power requirements.

The structure leverages the exceptional optical properties of thin-film materials, allowing light to be confined within deep subwavelength regions. This confinement results in significantly enhanced light-matter interaction, enabling efficient modulation of optical properties with minimal energy expenditure. Crucially, the modulation occurs within 10 nm-thick layers of ITO and TiO$_2$, a drastic reduction in active material volume compared to bulk modulation. This reduction in volume translates directly to lower power consumption. For the ITO/TiO$_2$/Al structure, the modulation of ITO can be achieved by applying a voltage between the Al and ITO layers. The energy consumption for this process can be estimated by modeling the ITO and TiO$_2$ layers as series capacitors: $U = \frac{1}{2}C_{t}V^2$, where $C_{t}$ is the total capacitance of the series combination of ITO and TiO$_2$ layers. For a ring resonator with an inner radius $R_{in} = 1300$ nm and an outer radius $R_{out} = 1500$ nm, and layer thicknesses of 10 nm for both ITO and TiO$_2$, the total capacitance is calculated. Assuming relative (static) permittivities of $\epsilon_{ITO} = 1.97$ and $\epsilon_{TiO2} = 2.43$, the total capacitance yields an energy consumption of approximately: $U \approx 3.39$ fJ for an applied voltage of $V = 2$ V. This ultra-low energy consumption is a direct consequence of the thin-film design and the enhanced field confinement within the subwavelength layers.
The integration of Sb$_2$S$_3$ as a PCM in the structure adds another layer of tunability. For a 40 nm-thick crystalline Sb$_2$S$_3$ film, it has been reported that a laser with 45 mW power and a 100 ms pulse duration can induce a phase change \cite{del}. In the proposed structure, the Sb$_2$S$_3$ layer is reduced to a thickness of 10 nm, significantly lowering the energy required for phase modulation. For the surface area defined by the ring resonator geometry, the energy required for phase change can be proportionally scaled down due to the reduced volume of PCM. Additionally, the inclusion of Al as a thermally conductive layer provides a means for localized heating of the PCM, either via resistive heating or optical excitation, ensuring efficient phase modulation. 

The proposed structure achieves high modulation speeds by utilizing deep subwavelength light confinement and ultrathin active layers with thicknesses of 10 nm. The modulation of the ITO layer under an applied voltage leverages its high carrier mobility and epsilon-near-zero (ENZ) properties to facilitate ultra-fast changes in optical characteristics—such as refractive index and absorption—on femtosecond timescales. Given the electron drift velocity across the 10 nm-thick ITO layer, the accumulation layer can theoretically form at frequencies exceeding the terahertz regime. However, when assessing the overall modulation speed, the capacitive RC-delay time likely becomes the dominant limiting factor, typically operating in the range of several hundred gigahertz \cite{sorger2012ultra,lin2015dynamically}. For the phase-change material (PCM), modulation speed can be enhanced by minimizing thermal mass (e.g., using thinner layers) and employing rapid localized heating techniques such as plasmonic heating or precise electrical pulsing. In the same class of low-loss PCMs, Sb$_2$Se$_3$ with a 30 nm thickness has demonstrated amorphization using a 9.5 V electrical pulse of 120 ns duration, while crystallization was achieved with a 4 V pulse of 20 µs duration. The switching energies for the amorphization and crystallization processes are 105 nJ and 1.2 $\mu$J, respectively\cite{yang2023non}.  Additionally, using a monolayer graphene heater, amorphization of a 30 nm Sb$_2$Se$_3$ layer has been demonstrated with a 6.8 V, 400 ns pulse width and an 8 ns trailing edge, whereas crystallization was achieved with a 4 V, 100 µs pulse width and a 120 µs trailing edge. The total switching energy for amorphization and crystallization is around 9.25 nJ and 1.28 $\mu$J, respectively \cite{fang2022ultra}. These results illustrate the potential for ultrafast and efficient modulation in the proposed structure.

In low-loss phase-change materials (PCMs), endurance is fundamentally a trade-off between switching speed and the number of cycles the material can withstand before degradation. Faster switching typically introduces greater thermal stress, reducing cycling lifetime. Nevertheless, by optimizing pulse conditions \cite{lawson2023optical}  and employing engineered laser pulse irradiation \cite{alam2024fast}, the endurance of these materials can exceed one million switching cycles. Additionally, encapsulating the PCM between Al$_2$O$_3$ layers helps mitigate degradation by reducing oxidation, thermal fatigue, and structural disorder \cite{fang2022ultra}. Furthermore, hybrid modulation with indium tin oxide (ITO) alleviates the load on the PCM by enabling amplitude control, thereby reducing the need for frequent phase transitions, which extends the overall device lifetime.

One of the system's most transformative features is its ability to operate near exceptional points (EPs), which amplify sensitivity and enhance functionality within a compact and energy-efficient framework. Through selective voltage application, the dimer supports at least 16 distinct states, offering high-dimensional programmability that significantly exceeds the capabilities of traditional photonic systems. By harnessing the 16 programmable states of our coupled resonator one can realize on chip optoelectronic logic gates, wherein each unique combination of four discrete amplitude levels and four phase settings corresponds to a distinct logic output, enabling multi valued logic operations within a single microring pair \cite{he2024chip}. In the quantum domain, mapping the 16 resonator states to distinct frequency bin superposition states enables on chip generation and coherent control of four dimensional qudits (D = 16), supporting high dimensional entanglement protocols for quantum communication \cite{kues2017chip}. 

Unlike conventional designs reliant on high-power nonlinear effects to modulate optical responses, this device achieves low-power modulation by exploiting the intrinsic properties of ENZ and PCM layers. This shift towards energy-efficient programmability represents a paradigm change for integrated photonics, addressing the critical need for sustainable and scalable technologies. Conventional thermo optic modulators, which rely on silicon’s thermo optic coefficient ($dn/dT\approx 1.8\times {{10}^{-4}}\ {{K}^{-1}}$), typically consume several milliwatts of continuous heating power. Likewise, ${{\chi }^{(3)}}$ based all optical Kerr modulators demand high intensity optical pumps, on the order of several hundred milliwatts, to induce refractive index changes on the order of $10^{-5}$. In stark contrast, ITO and PCM layers’ ultralow power consumption and phase-change material’s nonvolatility (once set, no further energy is required to maintain the state) enables a highly reconfigurable platform, multidimensional states with minimal overall power consumption.

The integration of PT-symmetric dynamics further enhances the system's tunability near EPs. This property not only expands the operational bandwidth but also improves control over phase and amplitude modulation, critical for applications in optical switching, signal processing, and reconfigurable photonic networks. Additionally, the hybrid plasmonic structure enables deep subwavelength confinement of light, dramatically increasing field intensities and ensuring high modulation efficiency. Such precise control over optical pathways, combined with minimal power consumption, is a key enabler for next-generation photonic devices.

A major strength of the proposed architecture lies in its compatibility with existing photonic integrated circuit (PIC) platforms. By overcoming challenges of dimensionality and scalability that have hindered earlier EP-based systems, this device demonstrates integrability with mature fabrication processes such as CMOS-compatible lithography. Since all deposition and annealing steps, including ITO ENZ layer and PCM, remain within the thermal budget ($<$ 400 °C) of CMOS back end of line, our device can be monolithically integrated alongside modulators, detectors, and multiplexers on a single silicon photonic chip, leveraging foundry proven design rules, layout libraries, and packaging workflows.This compatibility ensures that the system can be seamlessly incorporated into larger photonic networks, supporting applications such as neuromorphic computing, quantum photonic systems, and compact optical sensors. 

Beyond its immediate applications, the tunable photonic dimer establishes a versatile foundation for exploring multi-resonator configurations and advanced material platforms. For example, expanding the number of coupled resonators could enable even higher-dimensional state control, while novel material systems, such as active ENZ or hybrid PCM composites, could extend operational bandwidths. Moreover, incorporating AI-driven optimization techniques to dynamically adjust EP dynamics and material responses could unlock entirely new functionalities, such as real-time adaptive photonic networks.

This work paves the way for transformative innovations in integrated photonics, offering a path toward energy-efficient, highly tunable, and scalable optical systems. By addressing the critical demands of phase and amplitude modulation, deep subwavelength confinement, and compatibility with PICs, the tunable photonic dimer sets a new benchmark for photonic technology, driving progress in both fundamental research and practical applications.
\section{conclusion}
The proposed tunable photonic dimer represents a significant innovation in integrated photonics by uniquely combining Epsilon Near Zero (ENZ) and Phase-Change Materials (PCM) within a hybrid plasmonic architecture, achieving unparalleled control over light-matter interactions near Exceptional Points (EPs). This system's ability to simultaneously modulate phase and amplitude, leverage deep subwavelength light confinement, and dynamically program at least 16 distinct states through selective voltage control is unmatched in current photonic technologies. Its low-power operation, stemming from the intrinsic properties of ENZ and PCM, overcomes the high energy demands and limited tunability of traditional systems, while its compatibility with CMOS-compatible fabrication processes ensures seamless scalability and integration into photonic integrated circuits (PICs). This device establishes a new standard for programmable photonic platforms, empowering future innovations in reconfigurable optical networks, neuromorphic computing, quantum photonics, and compact on-chip sensing. Building on this work, researchers can explore multi-resonator configurations, novel materials, and AI-driven optimization techniques to create even more adaptive, intelligent, and high-performance photonic systems, pushing the boundaries of what is achievable in optical technologies.

\appendix

\section{}
To calculate transmission spectra of through port, we consider the coupled mode equation as
\begin{equation}
f\left( \begin{matrix}
   {{A}_{1}}  \\
   {{A}_{2}}  \\
\end{matrix} \right)=\left( \begin{matrix}
   {{f}_{1}}+i{{\gamma }_{1}} & \kappa   \\
   \kappa  & {{f}_{2}}+i{{\gamma }_{2}}  \\
\end{matrix} \right)\left( \begin{matrix}
   {{A}_{1}}  \\
   {{A}_{2}}  \\
\end{matrix} \right)+\left( \begin{matrix}
   i\sqrt{{{\gamma }_{c1}}}  \\
   0  \\
\end{matrix} \right){{S}_{+1}}
\end{equation}
, where $f$ is frequency of input source, ${{A}_{1}}$ and ${{A}_{2}}$ are field mode amplitudes, ${{S}_{+1}}$ is amplitude of input source, $\sqrt{{{\gamma }_{c1}}}$ is dissipation coupling of the source to the first ring. We may define the complex frequencies of the resonators as
\begin{equation}
{{\Omega }_{1}}={{f}_{1}}+i{{\gamma }_{1}}\quad ;\quad {{\Omega }_{2}}={{f}_{2}}+i{{\gamma }_{2}}
\end{equation}
Therefore, we can write eq. (1) as
\begin{equation}
f\left( \begin{matrix}
   {{A}_{1}}  \\
   {{A}_{2}}  \\
\end{matrix} \right)=\left( \begin{matrix}
   {{\Omega }_{1}} & \kappa   \\
   \kappa  & {{\Omega }_{2}}  \\
\end{matrix} \right)\left( \begin{matrix}
   {{A}_{1}}  \\
   {{A}_{2}}  \\
\end{matrix} \right)+i\left( \begin{matrix}
   \sqrt{{{\gamma }_{c1}}}  \\
   0  \\
\end{matrix} \right){{S}_{+1}}
\end{equation}
Solving the equation above for field amplitudes yields
\begin{align}
  & {{A}_{1}}=\frac{i\sqrt{{{\gamma }_{c1}}}(f-{{\Omega }_{2}})}{(f-{{\Omega }_{1}})(f-{{\Omega }_{2}})-{{\kappa }^{2}}}{{S}_{+1}} \\ 
 & {{A}_{2}}=\frac{i\sqrt{{{\gamma }_{c1}}}\kappa }{(f-{{\Omega }_{1}})(f-{{\Omega }_{2}})-{{\kappa }^{2}}}{{S}_{+1}}
\end{align}
From the field amplitudes we can calculate the transmission as
\begin{equation}
{{T}_{1\to 2}}={{\left| 1+\frac{\sqrt{{{\gamma }_{c1}}}{{A}_{1}}}{{{S}_{+1}}} \right|}^{2}}={{\left| 1+\frac{i{{\gamma }_{c1}}(f-{{\Omega }_{2}})}{(f-{{\Omega }_{1}})(f-{{\Omega }_{2}})-{{\kappa }^{2}}} \right|}^{2}}
\end{equation}
For the eigenvalue analysis, we consider the Hamiltonian of the system as
\begin{equation}
M=\left( \begin{matrix}
   {{\Omega }_{1}} & \kappa   \\
   \kappa  & {{\Omega }_{2}}  \\
\end{matrix} \right)
\end{equation}
, which have eigenvalues
\begin{equation}
\lambda_{1,2} =\frac{1}{2}\left( {{\Omega }_{1}}+{{\Omega }_{2}}\pm \sqrt{{{\left( {{\Omega }_{1}}-{{\Omega }_{2}} \right)}^{2}}+4{{\kappa }^{2}}} \right)
\end{equation}
To have exceptional point, we set 
\begin{equation}
{{\left( {{\Omega }_{1}}-{{\Omega }_{2}} \right)}^{2}}+4{{\kappa }^{2}}=0
\end{equation}
, which leads
\begin{equation}
{{f}_{1}}={{f}_{2}},\quad {{\gamma }_{2}}-{{\gamma }_{1}}=\pm 2\kappa
\end{equation}
\section{}
The electric potential within the ITO layer is described by Poisson’s equation:
\begin{equation}
\frac{{{d}^{2}}\phi (y)}{d{{y}^{2}}}=-\frac{\rho }{\varepsilon }=\frac{e(n(y)-{{n}_{0}})}{{{\varepsilon }_{0}}{{\varepsilon }_{ITO,s}}}
\end{equation}
, where $n(y)$ is electron density, ${{n}_{0}}$ is free carrier density, and ${{\varepsilon }_{ITO,s}}$ is static permittivity of ITO.
On the other hand, the density of states $g(E)$ for electrons in a 3D system is
\begin{equation}
g(E)=\frac{1}{2{{\pi }^{2}}}{{\left( \frac{2{{m}^{*}}}{{{\hbar }^{2}}} \right)}^{3/2}}\sqrt{E}
\end{equation}
, where ${{m}^{*}}$, $\hbar $ and $E$ is effective mass of the electrons, reduced Planck constant and energy relative to the conduction band edge.
The electron density $n(y)$ is obtained by integrating the density of states
\begin{align}
  & n(y)=\int_{0}^{{{E}_{F}}+e\phi (y)}{\frac{1}{2{{\pi }^{2}}}{{\left( \frac{2{{m}^{*}}}{{{\hbar }^{2}}} \right)}^{3/2}}\sqrt{E}}\ dE\\ \nonumber
 & \quad \quad =\frac{1}{3{{\pi }^{2}}}{{\left( \frac{2{{m}^{*}}}{{{\hbar }^{2}}} \right)}^{3/2}}{{\left( {{E}_{F}}+e\phi (y) \right)}^{3/2}} 
\end{align}
By replacing fermi energy level 
\begin{equation}
{{E}_{F}}=\frac{{{\hbar }^{2}}}{2{{m}^{*}}}{{\left[ 3{{\pi }^{2}}{{n}_{0}} \right]}^{2/3}}
\end{equation}
, in eq. (13), we can write
\begin{equation}
n(y)={{n}_{0}}{{\left( 1+e\phi (y)/{{E}_{F}} \right)}^{3/2}}
\end{equation}
Replacing the expression of $n(y)$ in eq. (B1) yields
\begin{equation}
\frac{{{d}^{2}}\phi (y)}{d{{y}^{2}}}=\frac{e{{n}_{0}}\left( {{\left( 1+e\phi (y)/{{E}_{F}} \right)}^{3/2}}-1 \right)}{{{\varepsilon }_{0}}{{\varepsilon }_{ITO,s}}}
\end{equation}
To solve the above equation, we define an intermittent function to convert second order differential equation to the first order as

\begin{align}
  & \phi (y)={{y}_{1}} \\ 
 & \frac{d\phi (y)}{dy}={{y}_{2}} 
\end{align}
Therefore, the set of equation
\begin{align}
  & {{{{y}'}}_{1}}={{y}_{2}} \\ 
 & {{{{y}'}}_{2}}=\frac{e{{n}_{0}}\left( {{\left( 1+e{{y}_{1}}/{{E}_{F}} \right)}^{3/2}}-1 \right)}{{{\varepsilon }_{0}}{{\varepsilon }_{ITO,s}}}
\end{align}
, can be solved numerically.

We consider boundary condition related to $\phi (y)={{y}_{1}}$ as
\begin{align}
  & {{y}_{1}}(0)=0 \\ 
 & {{y}_{1}}({{t}_{ITO}})={{V}_{ITO}}
\end{align}
, where ${{V}_{ITO}}$ can be calculated as following.
If we consider two layered structure of ITO and TiO$_2$ with permittivity ${{\varepsilon }_{ITO,s}}$ and ${{\varepsilon }_{TiO2}}$ and thickness of ${{t}_{ITO}}$ and ${{t}_{TiO2}}$ respectively, we may write capacitance of TiO$_2$, ITO and total capacitance of the structure as
\begin{align}
  & {{C}_{TiO2}}={{\varepsilon }_{0}}{{\varepsilon }_{TiO2}}\frac{A}{{{t}_{TiO2}}} \\ 
 & {{C}_{ITO}}={{\varepsilon }_{0}}{{\varepsilon }_{ITO,s}}\frac{A}{{{t}_{ITO}}} \\ 
 & {{C}_{t}}=\frac{A{{\varepsilon }_{0}}{{\varepsilon }_{TiO2}}{{\varepsilon }_{ITO,s}}}{{{\varepsilon }_{ITO,s}}{{t}_{TiO2}}+{{t}_{TiO2}}{{\varepsilon }_{ITO,s}}}
\end{align}
Therefore, we can write the total charge based on applied voltage $V$ as $Q=V{{C}_{t}}$
Therefore, 
\begin{equation}
{{V}_{ITO}}=\frac{Q}{{{C}_{ITO}}}=V\frac{{{\varepsilon }_{TiO2}}{{t}_{ITO}}}{{{\varepsilon }_{ITO,s}}{{t}_{TiO2}}+{{t}_{TiO2}}{{\varepsilon }_{ITO,s}}}
\end{equation}
\nocite{*}

\bibliography{sample}

\end{document}